\documentclass[twocolumn]{aastex631}
\usepackage{upgreek}
\usepackage{graphicx}
\usepackage{amsmath}
\usepackage{amssymb}
\usepackage{bm}
\usepackage[T1]{fontenc}
\usepackage{ae,aecompl}
\usepackage{color}
\usepackage[shortlabels]{enumitem}

\newcommand*{\bdv}[1]{\boldsymbol{\mathbf{#1}}}
\newcommand*{\uv}[1]{\hat{\boldsymbol{\mathbf{#1}}}}
\newcommand*{\rd}[2]{\frac{\mathrm{d}#1}{\mathrm{d}#2}}
\newcommand*{\p}[1]{\left(#1\right)}
\newcommand*{\s}[1]{\left[#1\right]}

\pdfoutput=1

\begin{document}

\title[Perpendicular Hot Jupiters]{High-Eccentricity Migration with Disk-Induced Spin-Orbit Misalignment: a Preference for Perpendicular Hot Jupiters}

\author{Michelle Vick}
\affiliation{Center for Interdisciplinary Exploration \& Research in Astrophysics (CIERA)\\ Northwestern University, Evanston, IL 60208, USA
}

\author{Yubo Su}
\affiliation{Department of Astrophysical Sciences \\ Princeton University, Princeton, NJ 08544, USA}

\author{Dong Lai}
\affiliation{Cornell Center for Astrophysics and Planetary Science, Department of Astronomy \\ Cornell University, Ithaca, NY 14853}


\begin{abstract}

High-eccentricity migration is a likely formation mechanism for many observed hot Jupiters, particularly those with a large misalignment between the stellar spin axis and orbital angular momentum axis of the planet. In one version of high-eccentricity migration, an inclined stellar companion excites von Zeipel-Lidov-Kozai (ZLK) eccentricity oscillations of a cold Jupiter, and tidal dissipation causes the planet's orbit to shrink and circularize. Throughout this process, the stellar spin can evolve chaotically, resulting in highly misaligned hot Jupiters. Previous population studies of this migration mechanism have assumed that the stellar spin is aligned with the planetary orbital angular momentum when the companion begins to induce ZLK oscillations.
However, in the presence of a binary companion, the star's obliquity may be significantly excited during the dissipation of its protoplanetary disk. We calculate the stellar obliquities produced in the protoplanetary disk phase and use these to perform an updated population synthesis of ZLK-driven high-eccentricity migration. We find that the resulting obliquity distribution of HJ systems is predominantly retrograde with a broad peak near 90$^\circ$. The distribution we obtain has intriguing similarities to the recently-observed preponderance of perpendicular planets close to their host stars.

\end{abstract}



\section{Introduction}\label{sec:Intro}

The existence of Hot Jupiters (HJs) is one of the oldest puzzles in exoplanet science. These giant planets with orbital periods $\lesssim 10$ days are surprising because the materials and conditions necessary to build gas giants do not exist so close to a host star. A variety of mechanisms have been suggested to explain HJ formation (for a review, see \citealp{Dawson18}). One promising theory is high-eccentricity migration, in which a fully-formed giant planet at a few AU is excited onto an eccentric orbit with a small pericenter distance. Over time, tidal dissipation within the planet due to the strong star-planet interaction at pericenter shrinks and circularizes the planet's orbit. One way to excite a giant planet onto an eccentric orbit relies on the von Zeipel-Lidov-Kozai (ZLK) effect \citep{Wu03,Fabrycky07,Naoz12,Correia12,Petrovich15a,Anderson16,Vick19a,Wang20,Rodet21}, in which a highly inclined stellar or planetary companion induces eccentricity oscillations in the planet's orbit, allowing the orbital eccentricity to climb to near-unity.

Historically, one strong point in favor of high-eccentricity migration is its ability to produce high obliquity HJ systems --- systems where the spin axis of the host star is highly misaligned with the orbital angular momentum of the planet. High obliquities have been observed in many HJ systems \citep{Hebrard08, Narita09,Winn09,Triaud10, Albrecht12,Winn15}. Such high obliquities are easier to explain via the dynamic process of high-eccentricity migration than through disk-driven migration, suggesting that at least a portion of the present-day HJ population formed through high-eccentricity migration \citep[e.g.][]{Rice22}.

ZLK-driven high eccentricity migration is especially efficient at generating highly misaligned systems. \citet{Storch14} revealed that spin-orbit coupling throughout ZLK cycles results in chaotic evolution of the stellar spin axis \citep[see also][]{Storch15,Storch17}. Population syntheses of ZLK high-eccentricity migration with a stellar companion showed that the resulting stellar obliquity distribution is bimodal, with peaks near 30-40$^\circ$ and 110-130$^\circ$ \citep{Fabrycky07,Correia12,Anderson16,Vick19a}; this bimodality can be understood as a bifurcation phenomenon of spin evolution during the high-eccentricity migration \citep{Storch17}. Note that ``retrograde" obliquities ($\theta_{\rm sl} > 90^\circ$) are not associated with any orbital flip of the planet.

Recently, \citet{Albrecht21} found that a few dozen misaligned HJ systems (and some Neptune-mass planets as well) have near-perpendicular stellar obliquities of $~80^\circ$-$125^\circ$. This range of obliquities falls directly in the valley of the bimodal distributions in \citet{Anderson16} and \citet{Vick19a}. One possibility is that the perpendicular planets began as retrograde and were guided toward perpendicular alignment through tidal dissipation in the host star \citep{Lai12, Rogers13, Anderson21}. But some of the perpendicular planets have orbits that are too wide for tides excited by the planet to be strong enough to drive efficient realignment. Most misaligned HJs orbit stars that do not have convective envelopes, and thus have very long realignment timescales. So there is an apparent tension between the ``preponderance of perpendicular planets'' and the obliquity distributions from previous studies of the ZLK high-eccentricity migration.

Many factors could affect the shape of the predicted obliquity distribution from ZLK high-eccentricity migration. A few of these, discussed in \citet{Anderson16}, include the stellar and planetary masses and the stellar rotation rate (dependent on the stellar type). For larger planet masses and more rapidly rotating stars, the resulting stellar obliquity distributions include perpendicular planets \citep[see Figure 25 of][]{Anderson16}. Most importantly, ``primordial'' misalignment (i.e. the stellar obliquity before the ZLK oscillation starts) can have a huge effect on the obliquities of resulting HJ systems \citep[Figure 26 of][]{Anderson16}. The vast majority of previous works on ZLK high-eccentricity HJ formation have assumed that the spin axis of the star is initially aligned with the orbital angular momentum of the planet. In reality, this is a special, even unlikely, case.

In the infancy of a giant planet, its orbit is strongly coupled to the protoplanetary disk. \citet{Batygin12} suggested that an inclined binary companion could change the orientation of a protoplanetary disk, generating a primordial misalignment between the stellar spin axis and the planet's orbit. Including the stellar spin-disk coupling in this scenario leads to more dramatic excitation of spin-orbit misalignment \citep{Batygin13, Lai14, Spalding14}. \citet{Zanazzi18} conducted a detailed modeling of the star-planet-disk-companion system and concluded that planetary systems with cold Jupiters (the starting point for high-eccentricity migration), but not HJs, could attain significant stellar obliquities as the protoplanetary disk dissipates.

Here, we develop a simple model for this ``disk+companion''-driven obliquity excitation. We then use these obliquities as the initial condition to generate the most realistic population synthesis to date of HJ formation via ZLK-driven high-eccentricity migration.
Our results reveal that primordial misalignment has a critically important effect on the predicted stellar obliquity distribution of HJs formed via this mechanism.

\section{Primordial Misalignment: Stellar Obliquity After Disk Dissipation} \label{sec:DiskDissipation}
\begin{figure*}
    \centering
    \includegraphics{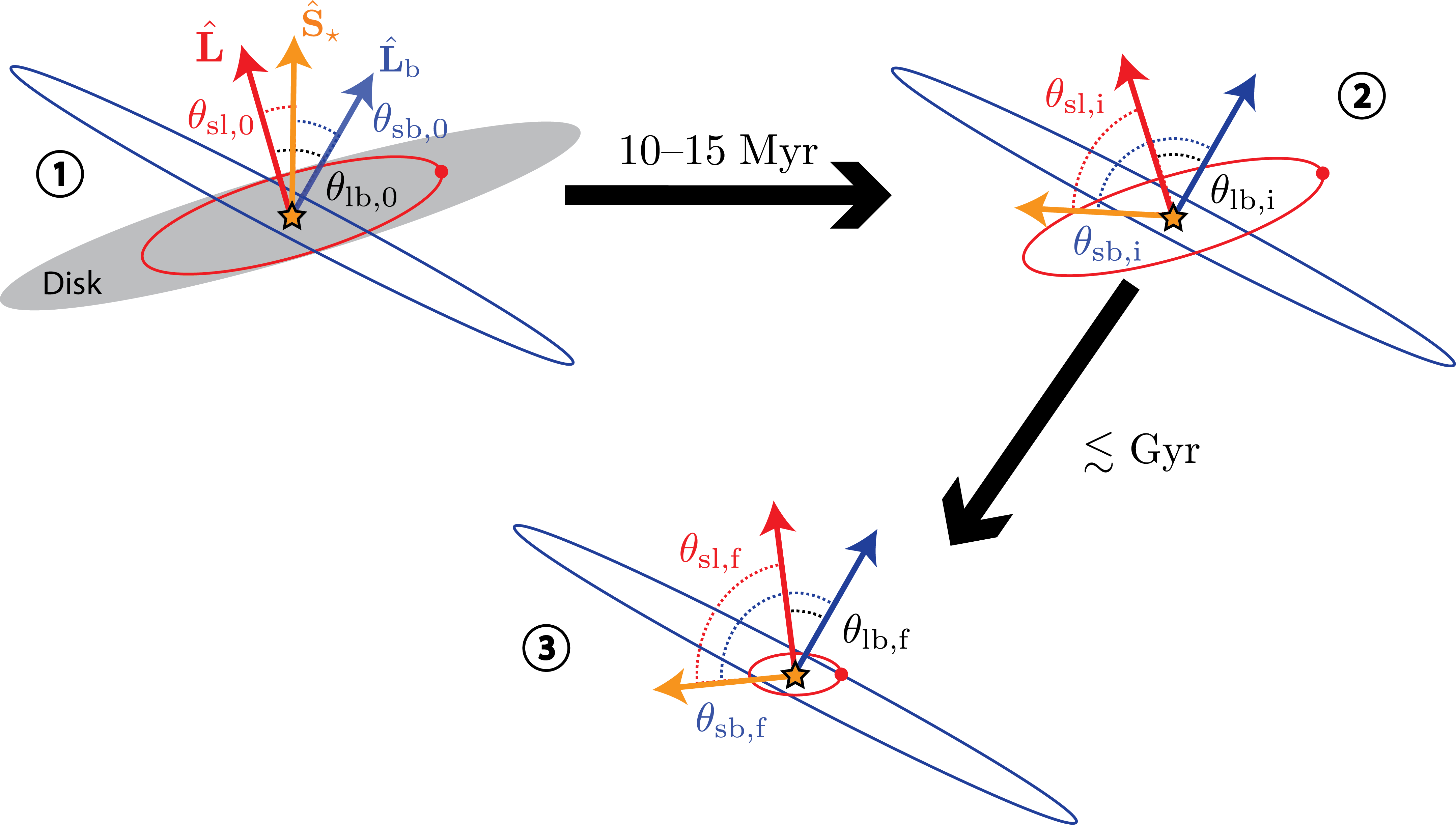}
    \caption{The proposed hot Jupiter (HJ) formation scenario. Initially, in panel (1), a proto-HJ (red) and a protoplanetary disk (grey) are both orbiting their host star, which also has a binary companion (blue).
    The stellar spin angular momentum and binary orbital angular momentum are denoted by $\uv{S}_{\star}$ and $\uv{L}_{\rm b}$ respectively.
    The disk and planet's angular momenta are assumed to be strongly coupled and evolve as a single angular momentum vector, $\uv{L}$.
    The mutual inclinations of the three vectors $\uv{S}_{\star}$, $\uv{L}$, and $\uv{L}_{\rm b}$ are described by the three angles $\theta_{\rm sb}$, $\theta_{\rm lb}$, and $\theta_{\rm sl}$ (the ``0'' subscript denotes the initial value of the angle). We assume $\theta_{\rm sl, 0} = 0$.
    In panel (2), after the protoplanetary disk has dissipated, the spin of the star changes orientation, and the relative angles among the angular momenta are notated with a ``i'' subscript (for their intermediate values).
    Finally, in panel (3), after high-eccentricity migration has resulted in a HJ, the relative angles among the angular momenta are denoted with a ``f'' subscript.
    Note that while $\theta_{\rm sb,f}$ and $\theta_{\rm sl,f}$ are approximately constant in time, $\theta_{\rm lb,f}$ can still undergo large oscillations.}\label{fig:cartoon}
\end{figure*}
It is known that the obliquity of a star can be significantly affected during the dissipation of its protoplanetary disk in the presence of a distant stellar binary companion \cite[e.g.][]{Batygin13, Lai14,Spalding14,Zanazzi18}. In this section, we briefly discuss the obliquity dynamics during this phase, and provide more detailed analytical results in Appendix~\ref{s:app_diskeom}.

\begin{figure*}
    \centering
    \includegraphics[width=5 in]{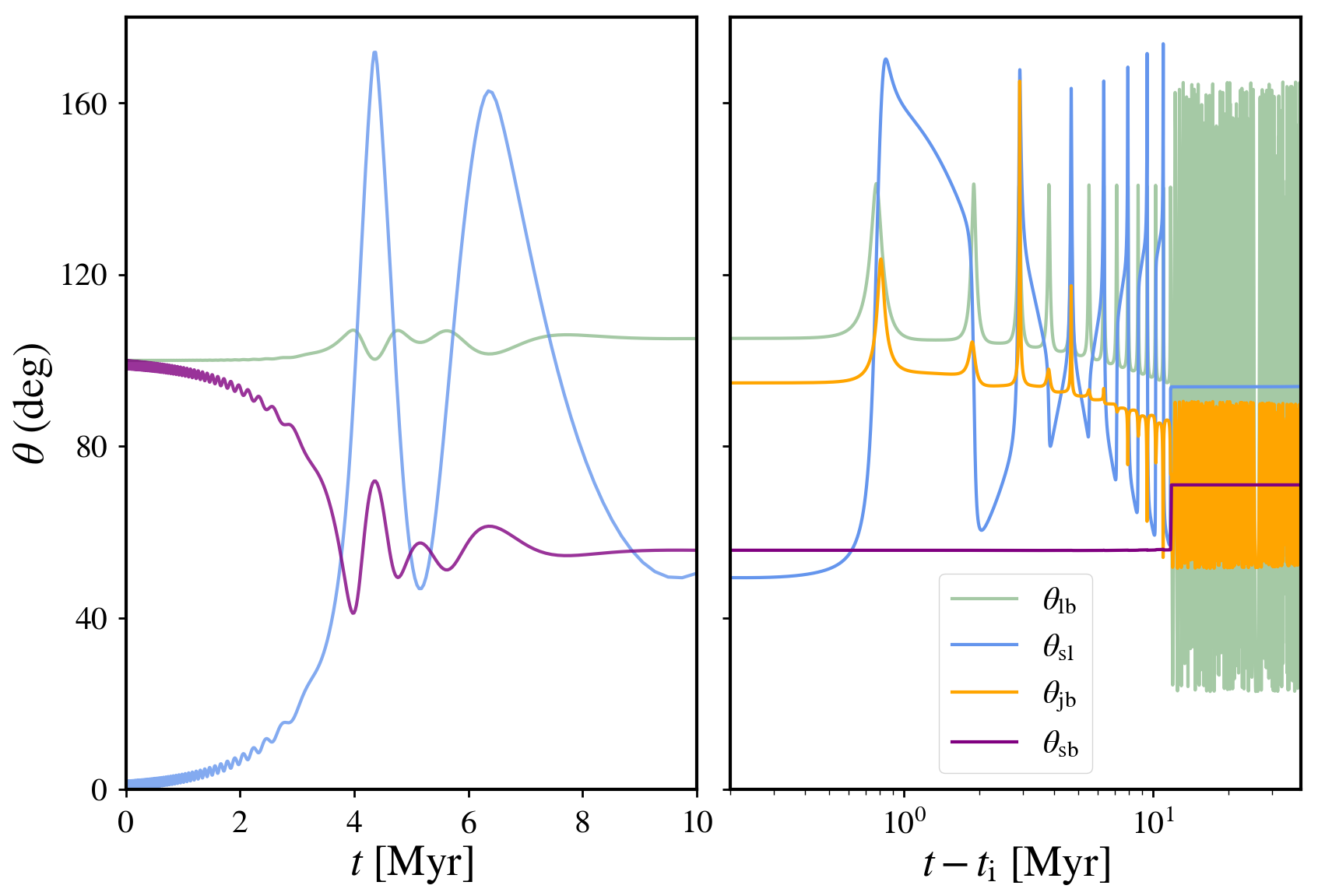}
    \caption{Left Panel: The evolution of various spin and orbit angles during the disk dissipation stage, obtained by integrating Equations~(\ref{eq:dsdt},~\ref{eq:dldt},~\ref{eq:md}). The initial conditions are $\theta_{\rm sl} = 0^\circ$ and $\theta_{\rm lb, 0} = \theta_{\rm sb, 0} = 100^\circ$. The integration is run for $10\tau_{\rm d} = 10~\mathrm{Myr}$, where $\tau_{\rm d} = \mathrm{Myr}$ is the characteristic disk lifetime.
    Right Panel: The continued evolution of the relevant spin and orbital angular momentum vectors throughout the migration of a giant planet. The plot is truncated before the planet's orbit circularizes fully. This panel also includes $\theta_{\rm jb}$, the angle between $\bdv{J} = \bdv{S}_\star + \bdv{L}$ and $\bdv{\textbf{L}}_{\rm b}$. The eccentricity of the binary companion is $e_{\rm b} = 0.6,$ and at the onset of ZLK oscillations ($t_{\rm i}$), $\theta_{\rm lb,i} = 105.1^\circ$ and $\theta_{\rm sl,i} = 50.43^\circ.$ At $t-t_{\rm i} \approx 10$~Myr, after multiple ZLK cycles, the orbital eccentricity is excited to near-unity, and chaotic dynamical tides rapidly shrink the orbit. Beyond this point, the orbital evolution of the planet is decoupled from the influence of the stellar companion, and the spin and orbit angles continue to behave in a similar way as the planetary orbit circularizes over another few 10s of Myrs.}
    \label{fig:AngleEvolution}
\end{figure*}

We assume that the cold Jupiter embedded in the protoplanetary disk is strongly coupled to the disk such that their angular momenta remain aligned. The configuration of the star-planet-disk-binary system is then specified by three unit angular momentum vectors: $\uv{S}_\star$, the spin axis of the star; $\uv{L}$, the shared angular momentum axis of the disk and planet; and $\uv{L}_{\rm b}$, the angular momentum axis of the binary. These three vectors precess about one another under their mutual gravitational torques. The relative orientations of these three vectors can be described using the three angles
\begin{align}
    \cos \theta_{\rm sb} &\equiv \uv{S}_\star \cdot \uv{L}_{\rm b},&
    \cos \theta_{\rm lb} &\equiv \uv{L} \cdot \uv{L}_{\rm b},&
    \cos \theta_{\rm sl} &\equiv \uv{S}_\star \cdot \uv{L}.
    \label{eq:def_I}
\end{align}
We denote the initial values of these three angles by $\theta_{\rm sb, 0}$, $\theta_{\rm lb, 0}$, and $\theta_{\rm sl, 0}$ respectively, and we denote the values once the protoplanetary disk has dissipated by $\theta_{\rm sb, i}$, $\theta_{\rm lb, i}$, and $\theta_{\rm sl, i}$. The latter notation indicates the ``intermediate'' values upon the end of disk dissipation but before ZLK-driven migration. For clarity, the definitions of these angles are also shown in Figure~\ref{fig:cartoon}. For the system's initial conditions, we always assume that the star and disk are initially aligned, so $\theta_{\rm sl, 0} = 0$ and $\theta_{\rm lb, 0} = \theta_{\rm sb, 0}$. An example of the system's disk-driven evolution when $\theta_{\rm lb, 0} = 100^\circ$ is shown in the left panel of Figure~\ref{fig:AngleEvolution}.

\begin{figure}
    \centering
    \includegraphics[width=3 in]{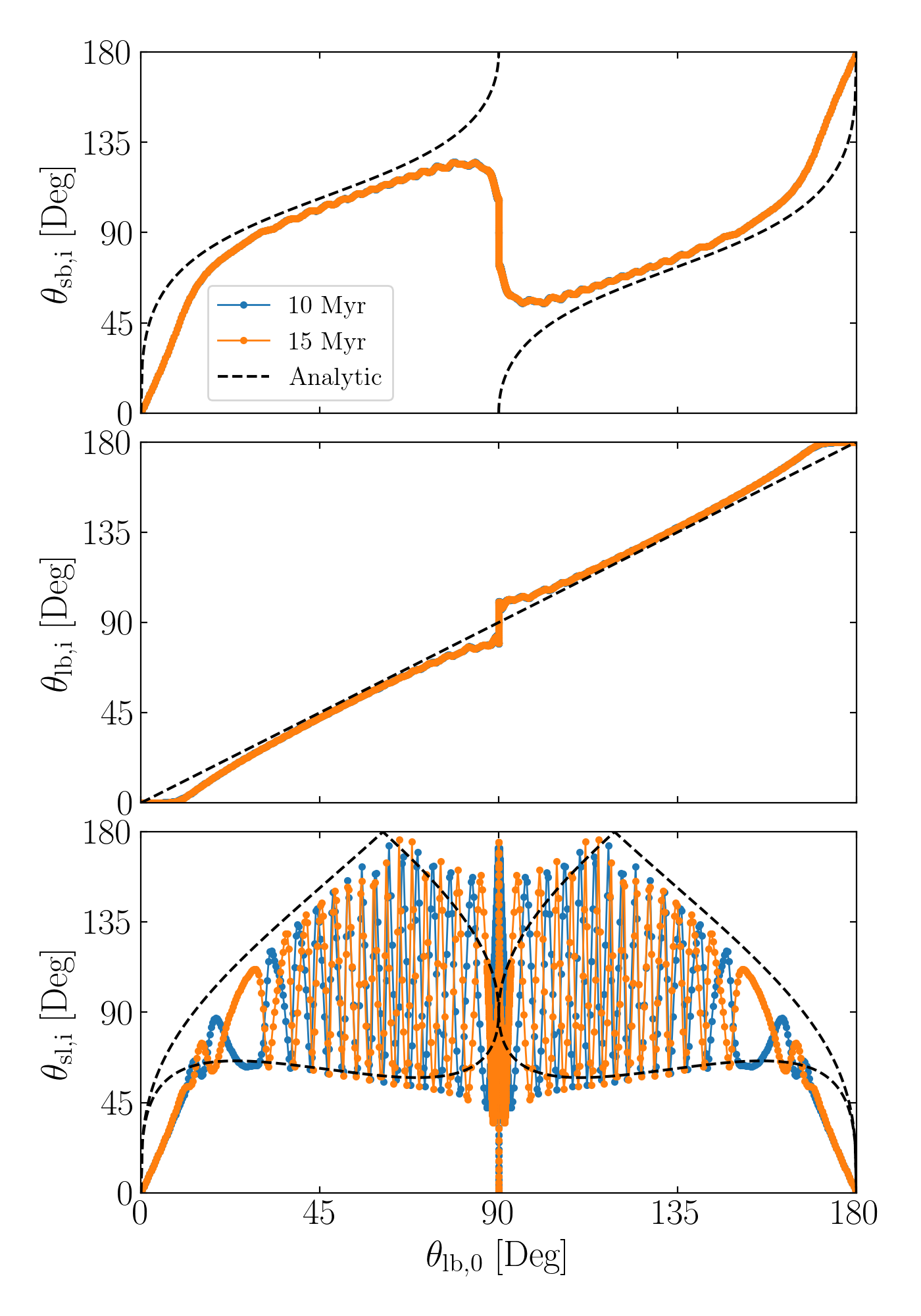}
    \caption{The intermediate angles $\theta_{\rm sb, i}$, $\theta_{\rm lb, i}$, and $\theta_{\rm sl, i}$ (recorded at the end of integrations such as those shown in the left panel of Figure~\ref{fig:AngleEvolution}) as a function of the initial binary inclination $\theta_{\rm lb, 0}$. The results for an integration time of either $10~\mathrm{Myr}$ or $15~\mathrm{Myr}$ are shown. The black dashed lines illustrate the approximate analytical model given by Equations~(\ref{eq:theta_sbi}--\ref{eq:theta_sli}). Good agreement is expected except when $\theta_{\rm lb, 0}$ is close to $0^\circ$, $90^\circ$, or $180^\circ$.}\label{fig:YS_thetas}
\end{figure}

We are interested in the relative orientations of the stellar spin and the planet after the disk has fully dissipated. For a schematic, see panel 2 of Figure~\ref{fig:cartoon}. Figure~\ref{fig:YS_thetas} shows $\theta_{\rm sb, i}$, $\theta_{\rm lb, i}$, and $\theta_{\rm sl, i}$ for all possible values of $\theta_{\rm lb, 0}$ when allowing the disk to dissipate for $t_{\rm f} = 10~\mathrm{Myr}$ and for $15~\mathrm{Myr}$; both disk lifetimes yield similar results.

A theoretical analysis of the system dynamics in the limit where the spin angular momentum of the star is negligible compared to that of the combined planet and disk reproduces the key features of Figure~\ref{fig:YS_thetas}. We relegate the details of this calculation to Appendix~\ref{ssec:approx_disk}, but its results can be summarized simply as:
\begin{align}
    \cos \theta_{\rm sb, i} &=
    \begin{cases}
        2\p{1 + \tan^{2/3} \theta_{\rm lb, 0}}^{-3/2} - 1
            & \theta_{\rm lb, 0} < 90\\
        1 - 2\s{1 + \tan^{2/3} (180 - \theta_{\rm lb, 0})}^{-3/2}
            & \theta_{\rm lb, 0} > 90
    \end{cases}\label{eq:theta_sbi},\\
    \theta_{\rm lb, i} &= \theta_{\rm lb, 0}\label{eq:theta_lbi}.
\end{align}
Finally, though $\theta_{\rm sl}$ varies rapidly later in the disk phase, a geometric analysis shows that $\theta_{\rm sl, i}$ is bounded by
\begin{equation}
    \theta_{\rm sl, i} \in
        \left[
            |\theta_{\rm sb, i} - \theta_{\rm lb, i}|,
            \min\Big(
            \theta_{\rm sb, i} + \theta_{\rm lb, i},
            360^\circ - (\theta_{\rm sb, i} + \theta_{\rm lb, i}) \Big)
        \right].\label{eq:theta_sli}
\end{equation}
This oscillation range is related to the rapid rate of change of the angle $\phi_{\rm sl}$, the angle between the projections of $\bdv{S}$ and $\bdv{L}$ onto $\bdv{L}_{\rm b}$. Note that $\phi_{\rm sl}$ advances uniformly (and rapidly) as the orbit precesses about the binary axis. Since (i) $\theta_{\rm sb}$, $\theta_{\rm lb}$, and $\phi_{\rm sl}$ together fully specify the mutual orientations of $\bdv{S}$, $\bdv{L}$, and $\bdv{L}_{\rm b}$, and (ii) $\theta_{\rm sb}$ and $\theta_{\rm lb}$ are approximately fixed when the disk mass is sufficiently small, we conclude that the specific final value of $\theta_{\rm sl, i}$ (within the range given by Equation~\ref{eq:theta_sli}) depends on the value of $\phi_{\rm sl, i}$. Since $\phi_{\rm sl}$ is rapidly and uniformly advancing, its final value $\phi_{\rm sl, i}$ is effectively randomly drawn from a uniform distribution over $[0, 2\uppi)$ depending on the final integration time.

Equations~(\ref{eq:theta_sbi}--\ref{eq:theta_sli}) are shown as the black dashed lines in each of the panels of Figure~\ref{fig:YS_thetas} respectively. When $\theta_{\rm lb, 0}$ is not too near any of $0^\circ$, $90^\circ$, or $180^\circ$, the agreement with numerical integrations is excellent. The rapid variation of $\theta_{\rm sl, i}$ can be seen in the densely-filled region in the bottom panel of Figure~\ref{fig:YS_thetas} as well as the difference between the results for the two different integration times (blue and orange curves).

We briefly discuss the origin of the disagreement of the numerical integrations with Equations~(\ref{eq:theta_sbi}--\ref{eq:theta_sli}). Note that the systems having $\theta_{\rm lb, 0} \approx 0^\circ$ or $\theta_{\rm lb, 0} \approx 180^\circ$ are not relevant for hot Jupiter formation; as such, we focus on the dynamics near $\theta_{\rm lb, 0} \approx 90^\circ$. As the disk dissipates and $\bdv{S}_\star$ becomes misaligned from $\bdv{L}_{\rm d}$, conservation of angular momentum necessitates a back-reaction torque on $\bdv{L}_{\rm d}$. This effect is generally small, as the resonant excitation of the obliquity occurs when $L_{\rm d} \gg S_\star$ and the back-reaction torque on the disk angular momentum is negligible. However, when $\theta_{\rm lb} \sim 90^\circ$, the disk-binary precession rate ($\omega_{\rm db} \cos \theta_{\rm lb}$) is slow, so resonance crossing (defined by the condition $\omega_{\rm sl} \sim \omega_{\rm lb} \cos \theta_{\rm lb}$) occurs when the disk mass is much lower. As such, the back-reaction torque on the disk's angular momentum can no longer be neglected when $\theta_{\rm lb, i} \simeq 90^\circ$, and $\theta_{\rm lb, i} \neq \theta_{\rm lb, 0}$, and planets are repelled from $\theta_{\rm lb,i}\sim90^\circ$.
Near these critical values of $\theta_{\rm lb, 0}$, the approximation that the stellar angular momentum is negligible is not valid. The resultant backreaction causes the distribution of $\theta_{\rm lb, i}$ to have a gap between $\sim 85^\circ$ and $\sim 95^\circ$ that can be seen in the middle panel of Figure~\ref{fig:YS_thetas}.

\section{Population Synthesis of ZLK-Driven Planet Migration}\label{sec:PopSynth}

Using the results of the previous section as initial conditions, we carry out a population synthesis of giant planets that undergo high-eccentricity migration due to ZLK eccentricity oscillations induced by a stellar companion.

\subsection{Model for ZLK Migration with Chaotic Tides}
We use the model developed in \citet{Vick19a} to couple the equations for ZLK migration with the chaotic evolution of the dynamical tide in the planet. Here, we provide a brief overview of the model.

During the high-eccentricity phase of the ZLK oscillations, the strong tidal forcing at pericenter excites oscillation modes in the planet. Over multiple orbits, the phases of these oscillations at pericenter determine the amount and direction of energy exchange between the orbit and the oscillation modes. If the planet's orbit has a small enough pericenter distance and a high enough eccentricity, the energy in the oscillation modes can grow chaotically over many orbits (\citealp{Vick18,Wu18,Vick19a, Yu21, Yu22}; see also \citealp{Mardling95a,Mardling95b,IP04,IP07,IP11}). When the modes reach a large enough amplitude, they dissipate non-linearly, making the tidal energy transfer irreversible, and driving the planet's orbit to decay and circularize. This process of ``chaotic tidal migration'' allows a highly eccentric cold Jupiter to become a (still eccentric) warm Jupiter on the timescale of $10^4-10^5$ yr, at which point weak tidal friction circularizes and shrinks the planet's orbit to that of a HJ on a much longer timescale of order Gyrs.

We consider the case where a distant binary star induces eccentricity oscillations in the orbital eccentricity of the planet. The timescale for quadrupole-order ZLK eccentricity oscillations is

\begin{align}
    t_{\rm ZLK} =& \left(\frac{10^6}{2 \uppi}~{\rm yr} \right) \left(\frac{M_b}{M_\odot }\right)^{-1}\left(\frac{M_\star}{M_\odot}\right)^{1/2}\left(\frac{a_{\rm p,0}}{1~{\rm au}}\right)^{-3/2} \nonumber \\
    &\times \left(\frac{a_{\rm b,eff}}{100~{\rm au}}\right)^3, \label{eq:tZLK}
\end{align}
where $M_{\rm b}$ is the mass of the stellar companion, $M_\star$ is the mass of the host star, $a_{\rm p,0}$ is the initial semimajor axis of the planet, and $a_{\rm b,eff} = a_{\rm b}(1-e_{\rm b}^2)^{1/2}$ is the effective semimajor axis of the companion with $a_{\rm b}$ and $e_{\rm b}$ the semimajor axis and eccentricity of the companion's orbit.

The model developed in \citet{Vick19a} evolves the $l=2, m=2$ f-mode of the planet (the oscillation that is most strongly excited by the stellar tidal potential) as well as changes to the orbital angular momentum, eccentricity and stellar spin due to the octupole-order ZLK effect. The importance of the octupole terms relative to the quadrupole terms is characterized by the parameter
\begin{equation}
    \epsilon_{\rm oct} = \frac{a_{\rm p}}{a_{\rm b}}\frac{e_{\rm b}}{1 - e_{\rm b}^2}. \label{eq:Octupole}
\end{equation}
At the start of an integration the planet's orbit is circular $e_0=0$, and the f-mode is not oscillating.

For a full description of the model, see section 3.1 of \citet{Vick19a} and references therein. One notable difference in this paper is that we focus our investigation on systems with an F-type host star. We therefore do not include stellar spin down due to magnetic braking, which is a much smaller effect for an F-type star than a G-type star over the timescale of HJ formation.

\subsection{Population Synthesis Setup and Methods}
Each integration in the population synthesis is for an F-type host star with $M_\star=1.4 M_\odot$, $R_\star= 1.4 R_\odot$, and a spin period of 3 days. The companion has $M_{\rm b}= M_\odot$. The planet model is a $\gamma=2$ polytrope with mass and radius $M_{\rm p}=M_{\rm J}$, $R_{\rm p}=1.6 R_{\rm J}$.

The orbit of the planet has initial semimajor axis $a_{\rm p,0} = a_{\rm p,i} = 5$~AU. We consider two choices of the semimajor axis of the companion, $a_{\rm b} = 150$~AU and $300$~AU. The population synthesis randomly samples $\cos(\theta_{\rm lb,0})=(-0.77,0.77)$ (the ZLK window), $e_{\rm b} = [0,0.8]$, and $\Omega_{\rm i} = [0,2 \uppi]$ with uniform probability, where $\Omega_{\rm i}$ is the longitude of the ascending node of the planet's orbit.

Each integration is stopped either at 10 Gyr or when one of the following criteria is met:
\begin{enumerate}[i)]
    \item If the threshold for chaotic tidal growth (see Equation 38 of \citealp{Vick19a}) is not met within $\min \left(150 t_{\rm ZLK}, 5 t_{\rm ZLK}/\epsilon_{\rm oct}\right)$, the calculation is terminated, and the system is labelled ``No Chaotic Migration.''
    \item If the system evolves for more than $2\times10^7$ planetary orbits before the f-mode energy reaches $0.1 G M_{\rm p}^2/R_{\rm p}$, the system is labelled as ``No Chaotic Migration'' and the integration is terminated. (see Section 3.1 of \citealp{Vick19a} or \citealp{Wu18} for discussion of this threshold).
    \item If the pericenter distance $r_{\rm p}$ of the planet's orbit dips below $2 R_{\rm p} (M_\star/M_{\rm p})^{1/3}$, the planet is considered to have been destroyed by tidal forces. The system is classified as ``Tidal Disruption'' and the integration is halted.
    \item Finally, if the planet's orbital eccentricity is circularized to $e=0.1$, the system is labelled as having undergone ``Chaotic Tidal Migration''.
\end{enumerate}
To increase the numerical efficiency of the population synthesis, we only include the effects of the dynamical tide, ZLK oscillations, and short range forces when they are physically important. See \citet{Vick19a} for a description of the conditions under which these effects are ``turned off'' in an integration.

At the end of each integration, if a HJ forms, we record the orientation angles of the system (see panel 3 of Figure~\ref{fig:cartoon}).

\section{Results}\label{sec:Results}

\begin{figure*}
    \centering
    \includegraphics[width=6in]{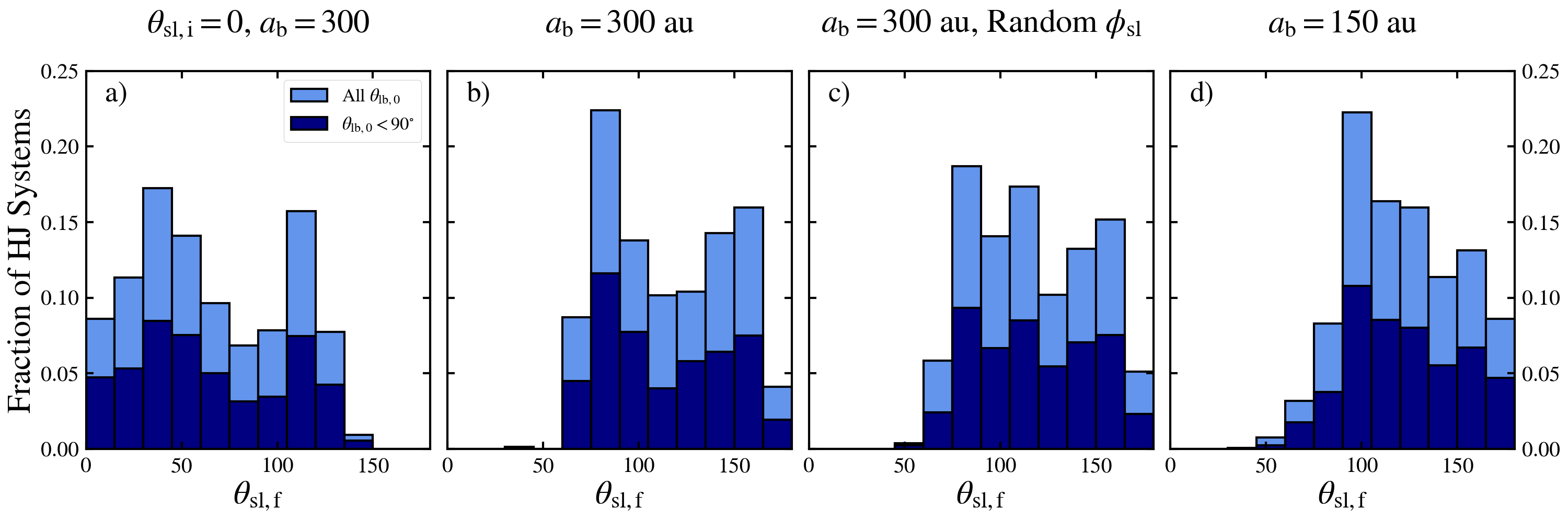}
    \caption{The distribution of stellar obliquities from the four sets of population syntheses described in Section~\ref{sec:PopSynth}. Panels (a)-(d) correspond to sets A-D. In all cases, the initial semimajor axis of the planet is $a_{\rm p,0}=5$~AU. In sets A-C $a_{\rm b}=300$~AU, while in set D, $a_{\rm b}=150$~AU. In set A, the stellar spin and orbital angular momentum begin aligned. In sets B, C, and D, $\theta_{\rm sb,i}$ and $\theta_{\rm lb,i}$ are determined by interpolation between the results of the integrations described in Section~\ref{sec:DiskDissipation}. In sets B and D, $\phi_{\rm sl,i}$ is also found by interpolation, while in set C, $\phi_{\rm sl,i}$ is randomly selected from [0, $2\uppi$). The distributions of $\theta_{\rm sl, f}$ for systems that are initially prograde, with mutual inclinations $\theta_{\rm lb,0} < 90^{\circ}$, are shown in dark blue.}
    \label{fig:SLDistribution}
\end{figure*}


\begin{figure*}
    \centering
    \includegraphics[width=6in]{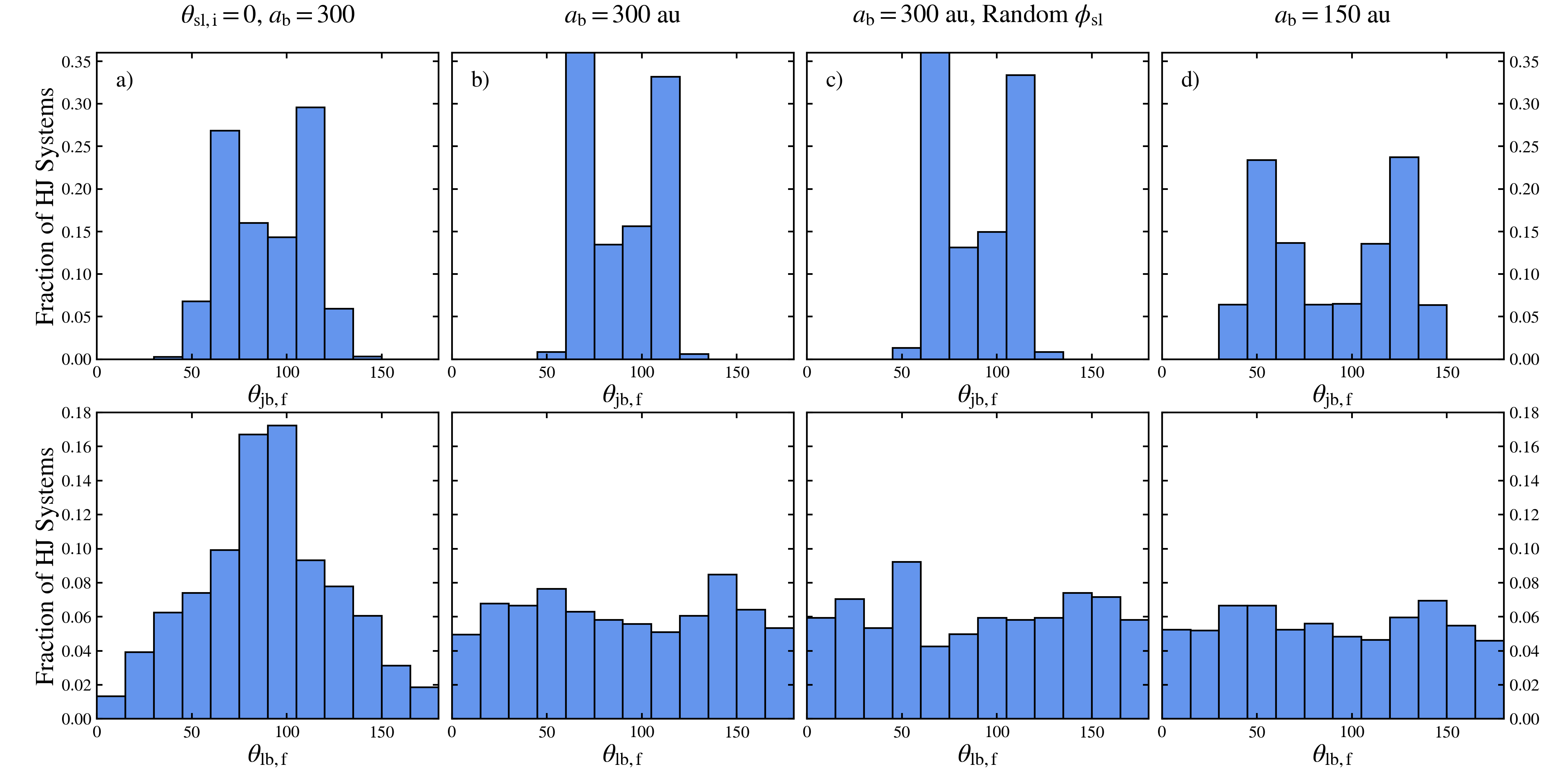}
    \caption{The distribution of $\theta_{\rm jb, f}$, the angle between $\bdv{J} = \bdv{S}_\star + \bdv{L}$ and $\bdv{L}_b$ (top row) and the distribution of $\theta_{\rm lb, f}$, the angle between $\bdv{L}$ and $\bdv{L}_b$ (bottom row). The columns are the same as in Figure~\ref{fig:SLDistribution}.}
    \label{fig:MutualInclination}
\end{figure*}

We ran four sets of population syntheses, A-D, each with $5\times 10^4$ systems. The initial parameters were selected as described in Section~\ref{sec:PopSynth}. In set A, $a_{\rm b}=300~$AU, and the stellar spin axis and orbital angular momentum axis begin aligned, i.e. $\theta_{\rm sl,i}=0$. This is the standard assumption adopted by previous studies. In sets B and C, $a_{\rm b}=300~$AU, and the intermediate $\theta_{\rm lb,i}$ and $\theta_{\rm sb,i}$ of each system are set by the outcome of the disk dissipation phase as described in Section~\ref{sec:DiskDissipation}. We interpolate between the data points shown in Figure~\ref{fig:YS_thetas} to obtain $\theta_{\rm lb, i}$ and $\theta_{\rm sb, i}$ for a given $\theta_{\rm lb, 0}$. To fully specify the initial orientations of the system angular momenta, we need to specify either $\theta_{\rm sl}$ or equivalently $\phi_{\rm sl}$ (defined in Section~\ref{sec:DiskDissipation}); we choose to use the angle $\phi_{\rm sl}$. In set B, we also use interpolation of the outcomes of the disk dissipation to obtain $\phi_{\rm sl}$, while in set C, we randomly sample $\phi_{\rm sl} = [0, 2\uppi)$. An example integration from set B is shown in the right panel of Figure~\ref{fig:AngleEvolution}. Lastly set D is the same as set B, but with $a_{\rm b}=150$~AU, and therefore an enhanced $\epsilon_{\rm oct}$ (see Equation~\ref{eq:Octupole}).

\subsection{HJ Stellar Obliquities}
    The final HJ stellar obliquities from the population syntheses are shown in Figure~\ref{fig:SLDistribution}. When the stellar spin and orbital angular momentum are aligned at the onset of ZLK cycles, the expected $\theta_{\rm sl,f}$ distribution is bimodal \citep{Storch14, Anderson16} with peaks near $40^\circ$ and $110^\circ$ [see panel a) of Figure~\ref{fig:SLDistribution}].

    When we account for the evolving orientation of the stellar spin as the disk dissipates, $\theta_{\rm sl,i}$ can adopt a broad range of values. Using the resulting $\theta_{\rm sl,i}$ from Section~\ref{sec:DiskDissipation}, we find that the distribution of HJ stellar obliquities is predominantly retrograde (see panels b,c, and d of Figure~\ref{fig:SLDistribution}). As these HJ systems continue to evolve, stellar tides will alter the ``final'' stellar obliquity $\theta_{\rm sl,f}$. In many cases, the dissipation of inertial waves in the star could act to move retrograde systems toward a $90^\circ$ misalignment \citep{Lai12}.

    We can understand the suppression of prograde $\theta_{\rm sl, f}$ as a combination of effects from both the disk and ZLK phases of evolution. In brief, the explanation has two pieces: (i) $\theta_{\rm sl, i}$ is preferentially retrograde (in fact, $\theta_{\rm sl, i} \gtrsim 50^\circ$) for HJ progenitors ($40^\circ \lesssim \theta_{\rm lb, 0} \lesssim 140^\circ$ in Figure~\ref{fig:YS_thetas}), and (ii) $\theta_{\rm sl, f}$ has a qualitatively similar distribution to $\theta_{\rm sl, i}$ (though for individual systems, $\theta_{\rm sl, f} \neq \theta_{\rm sl, i}$). We next justify each of these two claims individually.
    \begin{enumerate}[i)]
        \item At the end of the protoplanetary disk phase, Figure~\ref{fig:YS_thetas} shows that $\theta_{\rm lb, i} > 90^\circ$ whenever $\theta_{\rm sb, i} <  90^\circ$ (and vice versa) when restricting our attention to the systems that will become ZLK active, i.e.\ systems with $40^\circ \lesssim \theta_{\rm lb, i} \lesssim 140^\circ$. This is a consequence of Equation~\eqref{eq:qsbf}, which predicts $\theta_{\rm sb, i}$ exceeding $90^\circ$ when $\theta_{\rm lb, 0} \approx \theta_{\rm lb, i}$ is only $\sim 30^\circ$ for the given parameters. In fact, we only see $\theta_{\rm sl, i} \lesssim 50^\circ$ for $\theta_{\rm lb, 0}$ very near $90^\circ$. Thus, broadly speaking, $\theta_{\rm sl, i}$ is preferentially retrograde and is only rarely below $\sim 50^\circ$.
        
        \item To understand the evolution of $\theta_{\rm sl}$ during the ZLK phase, we instead study the evolution of the two angles $\theta_{\rm sb}$ and $\theta_{\rm lb}$. We also further subdivide the ZLK phase into the regimes of weak spin-orbit coupling (wide planet orbit) and strong spin-orbit coupling (close-in planet orbit).
        
        During the regime of weak spin-orbit coupling, the stellar spin is only weakly torqued by the planet and does not experience any other torques, so the spin orientation is roughly constant, and $\theta_{\rm sb}$ is fixed (see Figure~\ref{fig:AngleEvolution}). At the same time, while $\theta_{\rm lb}$ does oscillate, it does not cross $90^\circ$\footnote{It is a general feature of quadrupole-order ZLK oscillations that the orbit does not flip (the inclination does not cross $90^\circ$). While octupole-order ZLK can induce orbit flips in general, tidal precession suppresses these orbit flips \citep{liu2015suppression}.}. The above statement of ``$\theta_{\rm lb, i} > 90^\circ$ whenever $\theta_{\rm sb, i} <  90^\circ$ (and vice versa)'' continues to hold up to the onset of strong spin-orbit coupling. As such, prior to the onset of strong spin-orbit coupling, $\theta_{\rm sl}$ still cannot ever be smaller than $|\theta_{\rm lb} - \theta_{\rm sb}|$ (which remains strictly positive).

        Then, once strong spin-orbit coupling sets in, $\theta_{\rm sl}$ becomes approximately fixed, and equal to its value at the end of the weak-coupling phase discussed above. Our preceding argument shows that this value must be nonzero, and so $\theta_{\rm sl, f}$ has a lower bound (but no upper bound). According to our numerical results, this lower bound is in fact somewhat large ($\sim 50^\circ$). This results in a predominantly retrograde distribution.
    \end{enumerate}
    While this argument is not precise, it provides an accurate, qualitative justification for a remarkable (but not exclusive) preference for large $\theta_{\rm sl, f}$.

    The strong preference for retrograde stellar obliquities appears for both choices of the companion semimajor axis, $a_{\rm b}=150$ and $300~$AU. This suggests that this effect does not depend on the strength of the ZLK octupole effect, characterized by Equation~(\ref{eq:Octupole}).

    Unlike $\theta_{\rm lb,i}$ and $\theta_{\rm sb,i}$, $\phi_{\rm sl,i}$ varies significantly with small changes in $\theta_{\rm lb,0}$ and in the disk lifetime. For the robustness of our results, it is important to test whether
    small changes in these quantities (which are not well-constrained) have large effects on the expected HJ population. In population synthesis C, $\phi_{\rm sl}$ was randomly sampled with uniform probability from $[0,2\uppi)$. The resulting obliquity distribution in panel (c) of Figure~\ref{fig:SLDistribution} looks very similar to panel (b), which used $\phi_{\rm sl}$ taken from the outcome of the disk dissipation phase. This suggests that it is sufficient to use a randomly generated $\phi_{\rm sl}$ in future population syntheses, and that our results do not depend sensitively on assumptions about the disk lifetime.

    Recent observations suggest that systems with a retrograde mutual inclination ($\theta_{\rm lb,i} > 90^\circ$) are less common than prograde systems \citep{Dupuy22}. The dark blue histograms in Figure~\ref{fig:SLDistribution} show the results of the population synthesis for systems with $\theta_{\rm lb,i} < 90^\circ$. These look very similar in shape to the $\theta_{\rm sl,f}$ distributions when retrograde systems are included. 
    Even if systems with retrograde mutual inclinations are excluded from the population synthesis, the preference for retrograde obliquities in recently migrated HJs persists.

\subsection{Inclination of HJ Companions}\label{sec:MutualInclination}
    We also examine the expected inclination of HJ companions for HJs that form through high-eccentricity migration. The top row of Figure~\ref{fig:MutualInclination} shows the distributions of $\theta_{\rm  jb}$, the angle between $\textbf{J} = \bdv{S}_\star + \bdv{L}$ and $\bdv{L}_{\rm b}$, from our integrations. These distributions have peaks around $65^\circ$ and $115^\circ$. This feature has been seen in previous studies of ZLK high-eccentricity migration \citep{Vick19a}. When $\theta_{\rm sl,i}$ is calculated from a history of disk dissipation rather than fixed to 0, as in panels (b)-(d) of Figure~\ref{fig:MutualInclination}, these peaks are even more prominent.

    For an F-type host star, spin-down is negligible on the timescale of HJ formation, and $S_\star \lesssim L$ when the planet's orbit has circularized to $e=0.1$. At this point, the mutual inclination of the orbits $\theta_{\rm lb}$ oscillates rapidly relative to the timescale of orbital decay (see the right panel of Figure~\ref{fig:AngleEvolution}). 
    The $\theta_{\rm lb}$ distributions from our population syntheses (bottom panels in Figure~\ref{fig:MutualInclination}) sample a random phase of the $\theta_{\rm lb}$ oscillation for each system that successfully forms a HJ.

    The final $\theta_{\rm lb,f}$ distribution is very different for the $\theta_{\rm sl,i}=0$ case than for the cases with realistic $\theta_{\rm sl,i}$. If the stellar spin is aligned at the start of the ZLK cycles, the $\theta_{\rm lb,f}$ distribution has a peak at $90^\circ$, shown in panel (a) of Figure~\ref{fig:MutualInclination}. However, when $\theta_{\rm sl,i}$ is derived from a history of disk dissipation, the $\theta_{\rm lb,f}$ distribution is relatively flat with a slight dip at $90^{\circ}$, as for sets B-D shown in panels (b)-(d) of Figure~\ref{fig:MutualInclination}. This difference is due to the predominantly retrograde $\theta_{\rm sl,f}$ for systems in sets B-D. After the HJ orbit has circularized, the angles $\theta_{\rm sb,f}$ and $\theta_{\rm sl,f}$ are fixed but $\theta_{\rm lb,f}$ is still oscillating (see Figure~\ref{fig:AngleEvolution}). Therefore a large $\theta_{\rm sl,f}$ allows for a large range of mutual orbital inclinations. If $\theta_{\rm sl,f}$ tends closer to alignment, the distribution of mutual inclinations $\theta_{\rm lb,f}$ traces the distribution of $\theta_{\rm sb,f}$ more closely.

\subsection{HJ Formation Rate from ZLK High-eccentricity Migration}\label{ssec:ZLK_rate}

Disk dissipation leaves an imprint on the orientation of star-planet-companion systems. Notably,
not all values of $\theta_{\rm lb, i}$ can be attained (e.g.\ $\theta_{\rm lb, i} \approx 90^\circ$ is unattainable, see the middle panel of Figure~\ref{fig:YS_thetas}).
This can affect the predicted HJ formation rate via ZLK migration with chaotic tides. Although the population syntheses in this paper consider fixed values of $a$ and $a_{\rm b}$ and do not vary the mass and radius of the star or planet, they still provide some insight into how a primordial misalignment affects the HJ formation rate.

In general, highly misaligned planets (with $\theta_{\rm lb,i}$ close to $90^\circ$ are more likely to be driven by the ZLK effect to a small pericenter orbit, where disruption or HJ formation is possible. One might expect a gap around $\theta_{\rm lb,i}\sim 90^\circ$ to result in a lower HJ formation rate. Indeed, we find that our sets B and C have a reduced HJ formation rate compared to set A. In particular, the  HJ formation fraction for set A with $\theta_{\rm sl,i}=0$ is 23$\%$. For sets B and C, with $\theta_{\rm lb,i}$ and $\theta_{\rm sb,i}$ informed by the disk dissipation integrations, the formation rate is $13\%$. The HJ formation rate for population synthesis D (with $a_{\rm b}=150~$AU) is $26\%$. This increase could be due either to a slightly smaller gap in the $\theta_{\rm lb,i}$ distribution or due to the enhanced octupole effect (Equation~\ref{eq:Octupole}), which allows systems with a wider range of $\theta_{\rm lb,i}$ to reach very high eccentricities.

\section{Summary}

We have produced an updated population synthesis study of HJ formation via ZLK migration with a stellar companion. Previous studies of ZLK migration have assumed spin-orbit alignment ($\theta_{\rm sl,i}=0$) at the onset of ZLK oscillations. Under this assumption, the stellar obliquity distribution of the resulting HJ systems is bimodal with peaks near $30^\circ$-$40^\circ$ and $110^\circ$-$130^\circ$ and a paucity of polar planets. But $\theta_{\rm sl,i}=0$ is actually a special and unlikely case. Before ZLK oscillations take place, the giant planet is embedded in a protoplanetary disk, and the combined effects of companion-disk interaction, stellar spin-disk interaction, and disk dispersal give rise to a broad range of values for $\theta_{\rm sl,i}$ (see Figure~\ref{fig:YS_thetas}). When we incorporate these ``primordial'' obliquities into a population synthesis of ZLK migration (Section~\ref{sec:Results}), the predicted $\theta_{\rm sl,f}$ distribution of HJ systems from ZLK migration changes dramatically. We find that ZLK migration generates primarily retrograde stellar obliquities, with a broad peak around $90^\circ$ (see Figure~\ref{fig:SLDistribution}).
Over time, many of these host stars may evolve toward a perpendicular orientation due to stellar tides. Our result may therefore provide a possible explanation for the recently claimed ``preponderance of perpendicular planets'' among HJ systems \citep{Albrecht21}.


In addition, we show that ZLK-driven high-eccentricity migration results in a fairly flat distribution of $\theta_{\rm lb,f}$, the mutual inclination between the orbits of the planet and of the stellar companion. In contrast, when spin-orbit alignment is assumed at the onset of ZLK oscillations, the predicted $\theta_{\rm lb,f}$ distribution is peaked at $90^\circ$. Lastly, our results suggest that accounting for ``primordial" obliquities decreases the HJ formation fraction from this mechanism, but a larger scale population synthesis would be required to determine the size of the change.


\section{Acknowledgments}
We thank Justin Tan and Kassandra Anderson for helpful discussions.
This work has been supported
in part by the NSF grant AST-17152, the NASA grant
80NSSC19K0444, and the NASA FINESST grant 19-ASTRO19-0041. MV is supported by a Lindheimer Postdoctoral Fellowship at Northwestern University, and YS is supported by a Lyman Spitzer,
Jr. Postdoctoral Fellowship at Princeton University.



\appendix

\section{Disk Equations of Motion}\label{s:app_diskeom}

We consider a protostellar system consisting of a primary star with mass $M_{\star}$ surrounded by a planet with mass $m_{\rm p}$ embedded in a dissipating protoplanetary disk with mass $M_{\rm d}$ and an external binary companion with mass $M_{\rm b}$. The star is described by its radius $R_\star$, rotation rate $\Omega_{\rm \star}$, and spin angular momentum vector $\bdv{S}_\star = S_\star\uv{S}_\star$, where
\begin{equation}
    S_\star = k_\star M_\star R_\star^2 \Omega_\star,
\end{equation}
where $k_\star \simeq 0.2$ for a fully convective star. We further assume that the star has a rotation-induced quadrupole moment $J_2 M_\star R_\star^2$ with $J_2 = k_{\rm q\star} \Omega_\star^2 R_\star / (GM\star)$. We take the disk to have a surface density profile given by
\begin{equation}
    \Sigma = \Sigma_{\rm in} \frac{r_{\rm in}}{r},
\end{equation}
which extends from $r_{\rm in}$ to $r_{\rm out}$. Thus, the total disk mass is related to $\Sigma_{\rm in}$ by (assuming $r_{\rm out} \gg r_{\rm in}$)
\begin{equation}
    M_{\rm d} \simeq 2 \uppi \Sigma r_{\rm in} r_{\rm out}.
\end{equation}
The disk angular momentum vector is $\bdv{L}_{\rm d} = L_{\rm d} \uv{L}_{\rm d}$ with
\begin{equation}
    L_{\rm d} \simeq \frac{2}{3} M_{\rm d}\sqrt{GM_\star r_{\rm out}}.
\end{equation}
The planet has a circular orbit with radius $a_{\rm p}$. Throughout this paper, we assume that the planet's orbit axis $\uv{L}_{\rm p}$ is aligned with the disk axis $\uv{L}_{\rm d}$, i.e.\ $\uv{L}_{\rm p} = \uv{L}_{\rm d} \equiv \uv{L}$. The binary companion $M_{\rm b}$ has an orbital radius $a_{\rm b}$ which is at least a few times larger than $r_{\rm out}$. Since $L_{\rm b} \gg L$ and $S_\star$, we assume that $\uv{L}_{\rm b}$ remains fixed. Panel 1 of Figure~\ref{fig:cartoon} shows an illustration of the initial system and the relevant angular momentum vectors.

The system as described above has two dominant precessional effects: the mutual precession of the star and the combined planet-disk system, and the precession of the planet-disk system about the binary orbit. The spin vector $\uv{S}_\star$ evolves as
\begin{equation}
    \rd{\uv{S}_\star}{t} = -\omega_{\rm sl}
        (\uv{L} \cdot \uv{S}_\star)(\uv{L} \times \uv{S}_\star),
        \label{eq:dsdt}
\end{equation}
where $\omega_{\rm sl}$ is a combination of the spin-planet and spin-disk precession frequencies:
\begin{align}
    \omega_{\rm sl} \equiv{}& \omega_{\rm sd} + \omega_{\rm sp},\label{eq:def_wsl}\\
    \omega_{\rm sp} ={}& \frac{3k_{\rm q\star}}{2k_\star}
            \p{\frac{m_{\rm p}}{M_\star}}
            \p{\frac{R_\star}{a_{\rm p}}}^3 \Omega_\star\nonumber\\
        ={}& \frac{2\uppi}{2.2~\mathrm{Gyr}}
            \p{\frac{2k_{\rm q\star}}{k_\star}}
            \p{\frac{m_{\rm p}}{M_{\rm J}}}
            \p{\frac{a_{\rm p}}{5~\mathrm{AU}}}^{-3}\nonumber\\
        &\times \p{\frac{\Omega_\star}{0.1\Omega_{\rm \star, c}}}
            \p{\frac{R_\star}{2R_{\odot}}}^{3/2}
            \p{\frac{M_\star}{1.4M_{\odot}}}^{-1/2},\\
    \omega_{\rm sd} ={}& \frac{3k_{\rm q\star}}{4k_\star}
            \p{\frac{M_{\rm d}}{M_\star}}
            \p{\frac{R_\star^3}{r_{\rm in}^2r_{\rm out}}} \Omega_\star\nonumber\\
        ={}& \frac{2\uppi}{8.7~\mathrm{kyr}}
            \p{\frac{2k_{\rm q\star}}{k_\star}}
            \p{\frac{M_{\rm d}}{0.1M_\star}}
            \p{\frac{r_{\rm in}}{4 R_\star}}^{-2}\nonumber\\
        &\times \p{\frac{r_{\rm out}}{50~\mathrm{AU}}}^{-1}
            \p{\frac{\Omega_\star}{0.1\Omega_{\rm \star, c}}}
            \p{\frac{R_\star}{2 R_{\odot}}}^{-1/2}
            \p{\frac{M_\star}{1.4M_{\odot}}}^{1/2}.
\end{align}
Here, $\Omega_{\rm \star, c} \equiv \sqrt{GM_\star / R_\star^3}$ is the critical rotation rate of the star, and $M_{\rm J}$ is the mass of Jupiter. For the fiducial parameters, $0.1 \Omega_{\rm \star, c} = 2\uppi / \p{2.7~\mathrm{days}}$.

The disk and planet experience gravitational torques from both the oblate star and the binary companion. The joint disk and planetary axis $\uv{L}$ evolves according to
\begin{align}
    \rd{\uv{L}}{t} ={}&
        \omega_{\rm sl}\frac{S_\star}{L}
            \p{\uv{L} \cdot \uv{S}_\star} \p{\uv{L} \times \uv{S}_\star}
            \nonumber\\
        &- \omega_{\rm lb}
            \p{\uv{L} \cdot \uv{L}_{\rm b}}
            \p{\uv{L} \times \uv{L}_{\rm b}},\label{eq:dldt}
\end{align}
where $\omega_{\rm lb}$ is a combination of the planet-binary and disk-binary precession:
\begin{align}
    \omega_{\rm lb} \equiv{}& \omega_{\rm db}\frac{L_{\rm d}}{L}
            + \omega_{\rm pb}\frac{L_{\rm p}}{L},\label{eq:def_wlb}\\
    \omega_{\rm db} ={}& \frac{3M_{\rm b}}{8M_\star}
            \p{\frac{GM_\star r_{\rm out}^3}{a_{\rm b}^6}}^{1/2}\nonumber\\
        ={}& \frac{2\uppi}{0.17~\mathrm{Myr}}
            \p{\frac{M_{\rm b}}{M_\star}}
            \p{\frac{M_\star}{1.4M_{\odot}}}^{1/2}\nonumber\\
            &\times \p{\frac{r_{\rm out}}{50~\mathrm{AU}}}^{3/2}
            \p{\frac{a_{\rm b}}{300~\mathrm{AU}}}^{-3},\\
    \omega_{\rm pb} ={}& \frac{3M_{\rm b}}{4M_\star}
            \p{\frac{GM_\star a_{\rm p}^3}{a_{\rm b}^6}}^{1/2}\nonumber\\
        ={}& \frac{2\uppi}{2.8~\mathrm{Myr}}
            \p{\frac{M_{\rm b}}{M_\star}}
            \p{\frac{M_\star}{1.4M_{\odot}}}^{1/2}\nonumber\\
            &\times\p{\frac{a_{\rm p}}{5~\mathrm{AU}}}^{3/2}
            \p{\frac{a_{\rm b}}{300~\mathrm{AU}}}^{-3},
\end{align}
and $L \equiv L_{\rm p} + L_{\rm d}$ is the total angular momentum of the combined disk and planet. The angular momentum ratios are
\begin{align}
    \frac{S_\star}{L_{\rm d}}
        ={}& 0.003
            \p{\frac{k_\star}{0.2}}
            \p{\frac{M_{\rm d}}{0.1M_\star}}^{-1}
            \p{\frac{\Omega_\star}{0.1\Omega_{\rm \star, c}}} \nonumber\\
        &\times \p{\frac{R_\star}{2 R_{\odot}}}^{1/2}
            \p{\frac{r_{\rm out}}{50~\mathrm{AU}}}^{-1/2} ,\\
    \frac{S_\star}{L_{\rm p}}
        ={}& 2.7
            \p{\frac{k_\star}{0.2}}
            \p{\frac{m_{\rm p}}{M_{\rm J}}}^{-1}
            \p{\frac{M_\star}{M_\odot}}
            \p{\frac{\Omega_\star}{0.1\Omega_{\rm \star, c}}}\nonumber\\
        &\times
            \p{\frac{R_\star}{2R_{\odot}}}^{1/2}
            \p{\frac{a_{\rm p}}{5~\mathrm{AU}}}^{-1/2}.
\end{align}
We assume that the disk dissipates homologously, with its total mass evolving as:
\begin{equation}
    M_{\rm d}(t) = 0.1M_\star e^{-t / \tau_{\rm d}}. \label{eq:md}
\end{equation}
We take $\tau_{\rm d} = 1~\mathrm{Myr}$.

In summary, to model the evolution of the star-planet-disk-binary system, we numerically integrate Equations~\eqref{eq:dsdt} and~\eqref{eq:dldt} while the disk dissipates according to Equation~\eqref{eq:md}.

\subsection{Analytic Model: Heavy Disk Solution}\label{ssec:approx_disk}

While Equations~\eqref{eq:dsdt} and~\eqref{eq:dldt} are difficult to solve in general, the dynamics admit a simple approximate description. If $S_\star \ll L$ were satisfied all times, then the system would reduce to the so-called ``Colombo's Top'' model \citep{colombo1966, peale1969, peale1974possible, ward1975tidal, henrard1987}. Unfortunately, this condition is always violated once the disk has sufficiently dissipated, since then $L \approx L_{\rm p} \lesssim S_\star$. However, as long as this condition is well-satisfied throughout the secular resonance crossing (i.e.\ throughout the time when $\omega_{\rm sl} \sim \omega_{\rm lb}$) and is only violated once $\theta_{\rm sb}$ becomes roughly constant, then accurate predictions can be made for $\theta_{\rm sb, i}$. We briefly describe the obliquity excitation process as the protoplanetary disk dissipates including some new analytical results; this process is analogous to that described in \citet{2018AndersonTeeter}, where a detailed and complementary discussion of the dynamics can be found.

In the Colombo's Top model, the equilibria of the stellar obliquity are referred to as ``Cassini States'' (CSs). The number of CSs can be either two, when $\omega_{\rm lb} \gtrsim \omega_{\rm sl}$, or four, when $\omega_{\rm lb} \lesssim \omega_{\rm sl}$. When the disk is massive, $\omega_{\rm sl} \gg \omega_{\rm lb}$. In this regime, there is a CS that has nearly zero obliquity and is traditionally numbered CS1. Since initial spin-orbit alignment is assumed, the initial stellar spin very nearly occupies CS1. As the disk photoevaporates and $\omega_{\rm sl}$ decreases, the number of CSs changes from four to two. During this change, CS1 disappears \citep[due to a saddle-node bifurcation with CS4; see e.g.][]{henrard1987, 2018AndersonTeeter, su2020dynamics}, and the obliquity begins to oscillate with a large amplitude. At late times, the spin precession is much slower than the planet's orbital precession (about the binary axis), and so the spin instead precesses about the binary axis, which is the time-average of the 
planet's orbital angular momentum axis. If the evolution of the system is adiabatic, i.e.\ the disk photoevaporation is much slower than the all of the system's precession frequencies, then $\theta_{\rm sb, i}$ can be computed using the enclosed phase space area of the trajectory immediately after the disappearance of CS1 \citep[as first pointed out by][]{ward2004I}:
\begin{align}
    \cos \theta_{\rm sb, i} &=
    \begin{cases}
        2\p{1 + \tan^{2/3} \theta_{\rm lb, 0}}^{-3/2} - 1
            & \theta_{\rm lb, 0} < 90\\
        1 - 2\s{1 + \tan^{2/3} (180 - \theta_{\rm lb, 0})}^{-3/2}
            & \theta_{\rm lb, 0} > 90.
    \end{cases}\label{eq:qsbf}
\end{align}
This gives Equation~\eqref{eq:theta_sbi} in the main text.

Note that this expression is derived under the assumption that the spin angular momentum is negligible, and so it does not torque the combined disk and planet angular momentum. If this is true, then $\theta_{\rm lb}$ is a constant, or $\theta_{\rm lb, i} = \theta_{\rm lb, 0}$ (justifying Equation~\ref{eq:theta_lbi}). The middle panel of Figure~\ref{fig:YS_thetas} compares this result to the results of full numerical integrations; it can be seen that this is satisfied to good accuracy for a wide range of $\theta_{\rm lb, 0}$ except when $\theta_{\rm lb, 0}$ is close to
$0^\circ$, $180^\circ$, or $90^\circ$. These exceptions are due to angular momentum constraints. We omit discussion of the $\theta_{\rm lb, 0} \approx 0^\circ$ and $180^\circ$ exceptions in this paper since they do not result in planetary systems that undergo ZLK oscillations, and have discussed the deviation near $90^\circ$ in Section~\ref{sec:Results}. We discuss the case where $\theta_{\rm lb, 0}$ is close to $90^\circ$ in Section~\ref{ssec:ZLK_rate}. see Su et.\ al.\ (2022; in prep) for a more thorough discussion.

\bibliographystyle{aasjournal}
\bibliography{References,HJReferences}

\end{document}